\documentclass[aps,prl,twocolumn,showpacs,groupedaddress]{revtex4}
\usepackage{graphicx}
\usepackage{subfigure}
\usepackage{amsmath}
\usepackage{amssymb}
\usepackage{dsfont}
\usepackage{color}

\newcommand\tr{\mathrm{tr}}
\newcommand\down{\downarrow}
\newcommand\up{\uparrow}

\begin{document}

\title{Non-Abelian anyons: when Ising meets Fibonacci}

\author{E. Grosfeld$^{1}$ and K. Schoutens$^{2}$}
\affiliation{$^{1}$ Department of Physics, University of Illinois, 
1110 W. Green St., Urbana IL 61801-3080, U.S.A. \\
$^{2}$ Institute for Theoretical Physics, Valckenierstraat 65,
1018 XE Amsterdam, the Netherlands}

\date{13 Oct 2008, revised 23 Feb 2009}

\begin{abstract}
We consider an interface between two non-Abelian quantum Hall states: the 
Moore-Read state, supporting Ising anyons, and the $k=2$ 
non-Abelian spin-singlet state, supporting Fibonacci anyons. 
It is shown that the interface supports neutral excitations described by 
a 1+1 dimensional conformal field theory with central charge $c=7/10$. 
We discuss effects of the mismatch of the quantum statistical properties 
of the quasi-holes between the two sides, as reflected by 
the interface theory.
\end{abstract}


\maketitle

The quantum statistics of particles confined to two spatial dimensions is not
confined to be either bosonic or fermionic.  Particles called (Abelian) anyons 
pick up phase factors upon braiding, while for non-Abelian anyons braiding is
represented by non-trivial matrices acting on multi-component wavefunctions or
state vectors. Non-Abelian anyons offer most
exciting perspectives for what is called Topological Quantum Computation (TQC) 
\cite{Kitaev, TQCreview}.
The idea is that a collection of non-Abelian anyons, realized as excitations in a 
suitable quantum medium, open up a quantum register whose dimension depends 
on the number and the type of the anyons. This register can then be manipulated
via a braiding of world lines of the anyons, leading to quantum logic gates.

The leading candidate for physical systems that can support non-Abelian anyons 
are specific fractional quantum Hall (qH) liquids. Current experimental investigations seek
to confirm the tentative identification of the state underlying quantum Hall plateau
observed at filling fraction 5/2 with the Moore-Read (MR) state \cite{MooreRead}, 
or a close relative thereof \cite{levin,lee}. 
This state is known to support non-Abelian anyons of so-called Ising type,
the name deriving from an underlying algebraic structure which it has in common
with the 2D Ising model at criticality. The braid matrices for Ising anyons are 
non-trivial, but they fall short of allowing universal TQC. 

The other prototypical class of non-Abelian anyons are the so-called Fibonacci anyons. 
Their name derives from the fact that the dimensionality of the quantum register
for an $n$-anyon state is the $n$-th entry in the famous Fibonacci sequence 
$f_n=1,2,3,5,,..$, $f_n=f_{n-1}+f_{n-2}$. Matrices generated by successive braidings 
of such Fibonacci anyons are dense in the unitary group, implying that they
are universal for TQC . All logic operations on the quantum register can be 
approximated to arbitrary accuracy by successive braidings 
(see for example \cite{BonesteelHormoziZikosSimon}).

Two relatively simple quantum Hall states are known to support Fibonacci anyons (see, e.g. \cite{AS07}). 
The first is the so-called Read-Rezayi (RR) state with order $k=3$ clustering, at 
filling
$\nu=3/5$ (possibly related to a quantum Hall plateau observed at $\nu=12/5$). 
The other is the $k=2$ non-Abelian spin-singlet (NASS) state proposed by Ardonne 
and one of the present authors in 1999, at filling $\nu=4/7$. In many ways, this 
NASS state is similar to the MR state, the main difference being that it describes 
two species of fermions, which can be the spin-up and spin-down states of spin-1/2 electrons.  

For general quantum Hall liquids, an edge separating the liquid from vacuum carries
one or more gapless modes, described by a chiral conformal field theory (CFT). 
There is always a charge mode, which is responsible for the low-energy transport 
properties characteristic of quantum Hall
liquids. A non-Abelian state has neutral edge modes, which can be linked
to the fusion channel degeneracies of the bulk non-Abelian state. For the MR state 
this neutral mode is a Majorana (Ising) fermion (CFT with central charge $c=1/2$) while
for the $k=2$ NASS state the neutral modes are particular parafermions 
(descending from an $SU(3)$ structure, with a CFT central charge $c=6/5$), see  \cite{AS99,ARSS,AS07}  for details. 

In this letter we consider an interface between the MR liquid (supporting Ising 
anyons), and the $k=2$ NASS liquid (supporting Fibonacci anyons) and we 
investigate how the mismatch between the underlying topological orders plays 
out in the properties of this interface. We establish that the interface 
supports gapless neutral modes described by a specific CFT of 
central charge $c=7/10$. Dragging a Fibonacci anyon through this interface 
turns it into an Ising anyon, in 
the process exciting a specific ($h=3/80$) neutral interface mode. We also 
investigate 
to what extent processes where neutral bulk excitations tunnel to and from the
interface can relax the internal state of qubits spanned by pairs of 
quasi-holes. 

For a MR/NASS interface to be possible experimentally, it will be necessary
to have electronic interactions such both the MR state (in the polarized case)
and the NASS states (in the unpolarized case and at zero Zeeman splitting)
represent stable phases. Exact diagonalization studies \cite{Regnault09} in 
the second LL indicate that it is indeed possible to modify the 
Coulomb interaction such that both the MR and the NASS wavefunctions (for up 
to $N=12$ particles) have high overlaps with numerically obtained groundstates
in the appropriate regimes. 

\paragraph{MR and NASS states.}


The kinematic setting for the qH states we consider is the lowest Landau level (LLL), 
where $N$-body wavefunctions $\Psi(z_1,\ldots,z_N)$ factor into an analytic, 
polynomial expression $\tilde{\Psi}(z_1,\ldots,z_N)$ times a Gaussian factor. 
Below we first describe the bosonic versions of the MR and NASS wavefunctions (at 
filling $\nu=1$ and $\nu=4/3$, respectively). Their fermionic counterparts at 
$\nu=1/2$ and $\nu=4/7$ are obtained by multiplication with an overall Jastrow 
factor $\prod_{i<j}(z_i-z_j)$. 
The bosonic MR and NASS wavefunctions can be characterized
as the maximal density, zero energy eigenstates of  \cite{Greiter}
\begin{eqnarray}
H=\sum_{i<j<k} \delta^{(2)}(z_i-z_j)\delta^{(2)}(z_i-z_k).
\label{hamdeldel}
\end{eqnarray}
For the NASS states the coordinates $\{z_i\}$ split as $\{ z_i^\up, z_j^\down\} $.

The MR wavefunction can be written as
\begin{equation}
	\tilde{\Psi}_{\rm MR}=\frac{1}{\mathcal{N}}
            \sum_{S_1,S_2}\,
            \prod_{i<j \in S_1}(z_i-z_j)^2 \prod_{k<l \in S_2}(z_k-z_l)^2 \ ,
\end{equation}
where the sum is over all inequivalent ways of dividing the $N$ coordinates
into groups $S_1$, $S_2$ with $N/2$ coordinates each. In a similar way, the 
bosonic NASS wavefunction for $N_\downarrow$ spin-down particles and 
$N_\uparrow$ spin-up particles is
\begin{eqnarray}
	\tilde{\Psi}_{\mathrm{NASS}}=\frac{1}{\mathcal{N}}\sum_{S_1,S_2}\Psi^{221}_{S_1}(z^\uparrow_i,z^\downarrow_{j'})\Psi^{221}_{S_2}(z^\uparrow_k,z^\downarrow_{l'})\ ,
\end{eqnarray}
where the sum is over all inequivalent ways of dividing the coordinates to two groups, each containing $N_\uparrow/2$ spin-up and $N_\downarrow/2$ spin-down, and
\begin{eqnarray}	
\lefteqn{\Psi^{221}_{S_a}(z^\uparrow_i,z^\downarrow_{j'})=}
\\[2mm] &&
\prod_{i<j\in S_a}(z^\uparrow_i-z^\uparrow_j)^2
\prod_{i'<j'\in S_a}(z^\downarrow_{i'}-z^\downarrow_{j'})^2
\prod_{i,j'\in S_a}(z^\uparrow_i-z^\downarrow_{j'}) \ . \nonumber
\end{eqnarray}

\paragraph{Ising and Fibonacci anyons.}

For Ising anyons there are three particle types, $I$, $\psi$, and $\sigma$, with fusion rules
\begin{eqnarray}
	\psi\times\psi=I, \;\;\;\;\;
	\sigma\times\psi=\sigma, \;\;\;\;\;
	\sigma\times\sigma=I+\psi.
\end{eqnarray}
In addition, $I\times x=x$ for $x=I,\psi,\sigma$. For Fibonacci anyons there are only two particle types, $I$ and $\phi$,
\begin{eqnarray}
	I\times I=I, \;\;\;\;\; I\times \phi=\phi, \;\;\;\;\; \phi \times \phi=I+\phi.
\end{eqnarray}
The Virasoro primaries $I$, $\psi$ and $\sigma$ in the $c=1/2$ Ising CFT, of 
conformal dimensions $h_\psi=1/2$, $h_\sigma=1/16$, are in direct 
correspondence with the particle types $I$, $\psi$ and $\sigma$. The relation
between the Fibonacci particle types $I$ and $\phi$ and the $c=6/5$ parafermion theory 
is more subtle. The parafermion CFT has eight fields that are primary with respect to the 
parafermionic chiral algebra: the identity $I$, three $h=1/2$ parafermion fields 
$\psi_1$, $\psi_2$ and $\psi_{12}$, three $h=1/10$ spin fields $\sigma_\up$, 
$\sigma_\down$ and $\sigma_3$ and the $h=3/5$ spin field $\rho$. The correspondence is
\begin{equation}
I \leftrightarrow \{I,\psi_1,\psi_2,\psi_{12}\},
\qquad
\phi \leftrightarrow \{\sigma_\up,\sigma_\down,\sigma_{3},\rho\} \ .
\end{equation}
A further subtle point is that the parafermion sector denoted as `$\rho$'
contains two leading Virasoro primaries $\rho_c$ and $\rho_s$, of dimension 
$h=3/5$. The Virasoro fusion rule $\sigma_3 \sigma_3 = [ 1 + \rho_c]$
shows that $\rho_c$ acts as fusion channel changing operator for two
$\sigma_3$ fields, which correspond the spin-less quasi-holes over
NASS state. Similarly, the fusion rule $\sigma_\up \sigma_\down = \psi_{12} [1+ \rho_s]$ shows that $\rho_s$ changes the fusion channel for fields $\sigma_\up$ and
$\sigma_\down$, which come with spin-full quasi-holes. We refer to \cite{AS07}
for a complete description of the fusion rules and OPE's in the $c=6/5$ 
CFT.

\paragraph{Quasi-hole counting formulas and edge characters.}

Our strategy for obtaining the partition sum for a MR/NASS interface theory will be by
reduction from a counting formula for quasi-hole degeneracies in spherical geometry
[`giant hole approach'] .  In the presence of $N_\phi$ flux
quanta piercing through the sphere, the LLL orbitals form an angular momentum
multiplet with $L=N_\phi/2$, with, up to stereographical projection, the wavefunction 
$z^m$ corresponding to the orbital with $L_z=m-N_\phi/2$, for $m=0,\ldots,N_\phi$. 
The Hamiltonian eq. (\ref{hamdeldel}) acts on many-body wavefunctions with 
$N_\up$, $N_\down$ spin-up and spin-down electrons present. For $N_\up=N_\down$ and 
flux $N_\phi={3 \over 4}N-2$ there is a unique zero-energy  eigenstate, 
which is the bosonic NASS state whose asymptotic
filling is $\nu=4/3$. If we now add $\Delta N_\phi$ extra flux quanta and unbalance the 
numbers of up and down electrons, we create $n_\up$, $n_\down$ spin-up and spin-down
quasi-holes, with
$n_\up+n_\down = 4 \Delta N_\phi$,  $N_\downarrow+n_\down =  N_\uparrow+n_\up$.

The zero-energy quasi-hole states in the 
presence of $\Delta N_\phi$ are degenerate for two reasons. The first is a choice of orbital for the quasi-holes and the second is a choice of fusion channel.  The full structure of the space of zero-energy
states is captured by a zero-energy quasi-hole partition sum
$Z_{\rm sphere}[N_\up,N_\down;n_\up,n_\down](q)=\tr_{E=0} [q^{L_z}]$.
For the Laughlin and MR states, expressions for $Z_{\rm sphere}(q)$ have been given
in \cite{RR96}. For the $k=2$ NASS the following expression was obtained in \cite{ARSS}
 \begin{eqnarray}
    && \nonumber \sum_{\stackrel{\scriptstyle F_1\equiv N_\up \!\!\! \mod 2}{\scriptstyle F_2\equiv N_\down \!\!\! \mod 2}} q^{(F_1^2+F_2^2-F_1 F_2)/2}\left(\begin{array}{c}\frac{n_\uparrow+F_2}{2}\\F_1\end{array}\right)_q\left(\begin{array}{c}\frac{n_\downarrow+F_1}{2}\\F_2\end{array}\right)_q\\
    &&\quad\quad\quad\times\left(\begin{array}{c}\frac{N_\uparrow-F_1}{2}+n_\uparrow\\n_\uparrow\end{array}\right)_q\left(\begin{array}{c}\frac{N_\downarrow-F_2}{2}+n_\downarrow\\n_\downarrow\end{array}\right)_q.
    \label{qhNASS}
\end{eqnarray}
The $q$-binomial is (here $(q)_n\equiv (1-q)(1-q^2)\ldots(1-q^n)$)
\begin{eqnarray}
    \left(\begin{array}{c}n\\m\end{array}\right)_q\equiv\frac{(q)_n}{(q)_m(q)_{n-m}}.
\end{eqnarray}  

Putting $N_\up=N$, $N_\down=0$, $n_\up=n$, $n_\down=N+n$, the formula reduces to
the case of the $N$-particle MR state with $n$ quasi-holes,
\begin{eqnarray}
	    \label{eq:MR-counting}
    \sum_{F,(-1)^F=(-1)^N} q^{F^2/2}\left(\begin{array}{c}n/2\\F\end{array}\right)_q\left(\begin{array}{c}\frac{N-F}{2}+n\\n\end{array}\right)_q.
\end{eqnarray}


Let us now consider the MR state and demonstrate how to extract the edge characters from the bulk counting formula. The exact physical mechanism will be described in the next section, but here we notice that if we take a large number $n$ of quasi-holes in Eq.~(\ref{eq:MR-counting})
and take the limit $N\to\infty$, the counting formula reduces to  
\begin{eqnarray}
    \left[\frac{1}{(q)_n}\right]\left[\sum_{F=0,2,\ldots} q^{F^2/2}\left(\begin{array}{c}n/2\\F\end{array}\right)_q
    \right].
\end{eqnarray}
The first bracket coincides with the boson character when the limit $n\to\infty$ is 
taken. The second bracket coincides with the Ising
vacuum character when $n\to\infty$. The edge content of the MR state is thus completely 
reproduced.

\begin{table}
	\vspace{0.5cm}
    \begin{tabular}{|c|c||c|c|c|c|c|}
        \hline
         & h & $\epsilon$ & $\epsilon'$ & $\epsilon''$ & $\tilde{\sigma}$ & $\tilde{\sigma}'$\\
        \hline
        \hline
        $\epsilon$ & $1/10$ & $I+\epsilon'$ & & & & \\
        $\epsilon'$ & $3/5$ & $\epsilon+\epsilon''$ & $I+\epsilon'$ & & & \\
        $\epsilon''$ & $3/2$ & $\epsilon'$ & $\epsilon$ & $I$ & & \\
        $\tilde{\sigma}$ & $3/80$ & $\tilde{\sigma}+\tilde{\sigma}'$ & $\tilde{\sigma}+\tilde{\sigma}'$ & $\tilde{\sigma}$ & $I+\epsilon+\epsilon'+\epsilon''$ & \\
        $\tilde{\sigma}'$ & $7/16$ & $\tilde{\sigma}$ & $\tilde{\sigma}$ & $\tilde{\sigma}'$ & $\epsilon+\epsilon'$ & $I+\epsilon''$ \\
        \hline
	\end{tabular}
	\caption{Primary fields of the conformal field theory at central charge $c=7/10$, along with their conformal dimensions $h$, and fusion rules (see, e.g. \cite{CFTbook}).}
\label{cfttable}
\end{table}

\paragraph{MR/NASS interface.}
We now use the 'giant hole' technique to investigate the MR/NASS interface. 
We wish to consider a 2-fluid configuration on the sphere with  $N_{NASS}$ particles making 
up a NASS state and $N_{MR}$ particles making up a MR state, so that 
$N_\uparrow=\frac{1}{2}N_{NASS}+N_{MR}$ and $N_\downarrow=\frac{1}{2}N_{NASS}$.  
The number of flux quanta needed to accommodate this 2-fluid state is 
$N_\phi=\frac{3}{4}N_{NASS}+N_{MR}-2$. Comparing with a situation where all 
$N_{NASS}+N_{MR}$ particles form a NASS state we have an excess flux of
 $\Delta N_\phi=\frac{1}{4}N_{MR}$, giving rise
to the presence of $n_\uparrow+n_\downarrow=4 \Delta N_\phi=N_{MR}$ quasi-particles.
Using $N_\downarrow+n_\down =  N_\uparrow+n_\up$ we infer that $n_\uparrow=0$, 
$n_\downarrow=N_{MR}$. 

For the values of $N_\uparrow$, $N_\downarrow$, $n_\uparrow$ and $n_\downarrow$ thus
specified the Hamiltonian (\ref{hamdeldel}) allows a large number of zero-energy eigenstates,
as given in eq.~(\ref{qhNASS}). However, in the presence of more realistic Coulomb interactions
these states will no longer be degenerate. One expects that the lowest energy states will 
be phase-separated, with regions of NASS and MR liquids separated by an interface. The other
states in eq.~(\ref{qhNASS}) then correspond to excitations of this interface. One  can further stabilize 
such a configuration by assuming an orbital-dependent Zeeman term which favors the liquid to 
be spin-polarized in a specific region, say near the south pole on the sphere.  

In the limit of $N_\uparrow, \,N_\downarrow \to \infty$ eq.~(\ref{qhNASS}) reduces to a charge boson factor times the following factor, accounting for neutral interface excitations 
\begin{eqnarray}
    \sum_{F_1,F_2=0,2,4,\ldots} q^{(F_1^2+F_2^2-F_1 F_2)/2}\left(\begin{array}{c}\frac{n_\downarrow+F_1}{2}\\F_2\end{array}\right)_q\left(\begin{array}{c}F_2/2\\F_1\end{array}\right)_q.
\end{eqnarray}
This expression coincides with a finitized chiral character for the vacuum sector 
in a $c=7/10$ minimal model of CFT \cite{Feverati}. We conclude that the MR/NASS
interface supports neutral excitations described by this precise CFT [table
\ref{cfttable}].
\footnote{We have identified the interface excitations in a physical picture where the 2-fluid 
configuration is generated from a NASS state by accumulating spin-down quasi-holes near 
the south pole on the sphere. A possible strategy towards experimental realization of a MR/ NASS interface will be rather the opposite: start from the polarized MR state, reduce Zeeman splitting by applying hydrostatic pressure and then increase the filling, so as to nucleate islands of the NASS phase 
in the MR background.}. 

The fields of the CFT at $c=6/5$ can be written as a direct product of fields of the 
CFTs at $c=7/10$ and $c=1/2$. We identify the correspondence by the use of a 
character formula and through the discrete symmetries associated with the fields. 
This requires to consider an extended algebra produced
by explicitly adding a fermion parity operator to both the $c=7/10$ and the $c=1/2$ theories, $(-1)^{F}$ and $(-1)^{F'}$, which satisfy $\{(-1)^{F},\epsilon''\}=0$ and $\{(-1)^{F'},\psi\}=0$. The Ramond sector is then effectively 'doubled', so $\sigma$, $\tilde{\sigma}$ and $\tilde{\sigma}'$ are replaced respectively by $\sigma_\pm$, $\tilde{\sigma}_\pm$, and $\tilde{\sigma}'_\pm$, each having a well defined fermion parity given by the subscript. Their fusion rules are now constrained so that fermion parity is respected. 

The fields are related through the following relations
\begin{eqnarray}
    \nonumber && \sigma_{\downarrow}=\tilde{\sigma}_+\otimes \sigma_+ + \tilde{\sigma}_-\otimes \sigma_-,  \;\;\; \sigma_{\uparrow}=\tilde{\sigma}_+\otimes \sigma_- + \tilde{\sigma}_-\otimes \sigma_+,\\
    \nonumber&&\sigma_3=\epsilon'\otimes \psi+\epsilon\otimes I,\\
    \nonumber&& \psi_{1}=\tilde{\sigma}'_+\otimes \sigma_-+\tilde{\sigma}'_-\otimes \sigma_+, \;\;\; 	\psi_{2}=\tilde{\sigma}'_+\otimes \sigma_++\tilde{\sigma}'_-\otimes \sigma_-,\\
    \nonumber&&\psi_{12}=\epsilon''\otimes I+I\otimes \psi,\\
    &&\rho=\epsilon'\otimes I+\epsilon\otimes \psi,  \quad\quad\quad\quad I=\epsilon''\otimes \psi+I\otimes I.
    \label{eq:identifications}
\end{eqnarray}
where the notation $\phi_1\otimes \phi_2$ describes a direct product of a field in the $c=7/10$ theory ($\phi_1$) and a field in the $c=1/2$ theory ($\phi_2$). In addition, the two Virasoro primaries $\rho_s$ and $\rho_c$ have the following decompositions
\begin{eqnarray}
    \rho_s=\epsilon\otimes \psi, \;\;\;\;\; \rho_c=\epsilon'\otimes I.
    \label{eq:identifications-rho}
\end{eqnarray}

A physical way of viewing the creation of the $c=7/10$ edge between the MR state 
and the NASS state is by starting with counter-propagating edges, with $\bar{c}=1/2$ 
and $c=6/5=1/2+7/10$ respectively, and introducing tunneling between the two edges. 
As the tunneling increases, the counter-propagating Majorana fermions gap out, 
leaving behind only the $c=7/10$ edge. It is useful in this case to consider an inverted form of eq. (\ref{eq:identifications}), which contains explicitly both counter-propagating modes. In this way, one can identify those degrees of freedom which gap out and those which remain behind. For example, at the level of the characters the following relation holds
\begin{eqnarray}
	\nonumber &&\bar{\psi}\psi_{12}+\bar{I}_{1/2} I_{6/5}=\\
	&&\quad I_{7/10}(\bar{\psi}\psi+\bar{I}_{1/2}I_{1/2})+\epsilon''(\bar{\psi}I_{1/2}+\bar{I}_{1/2}\psi).
\end{eqnarray}
Here  $I_{1/2}$, $I_{7/10}$, and $I_{6/5}$ are the identity fields for the three theories. The combinations of fields appearing within brackets gap out when an effective mass term $m\bar{\psi}\psi$ is generated by tunneling, leaving only the $c=7/10$ degrees of freedom behind on the edge.

\begin{figure}[h]
  \hfill
  \begin{minipage}[t]{0.5\textwidth}
    \begin{center}
      \includegraphics[width=0.45\textwidth]{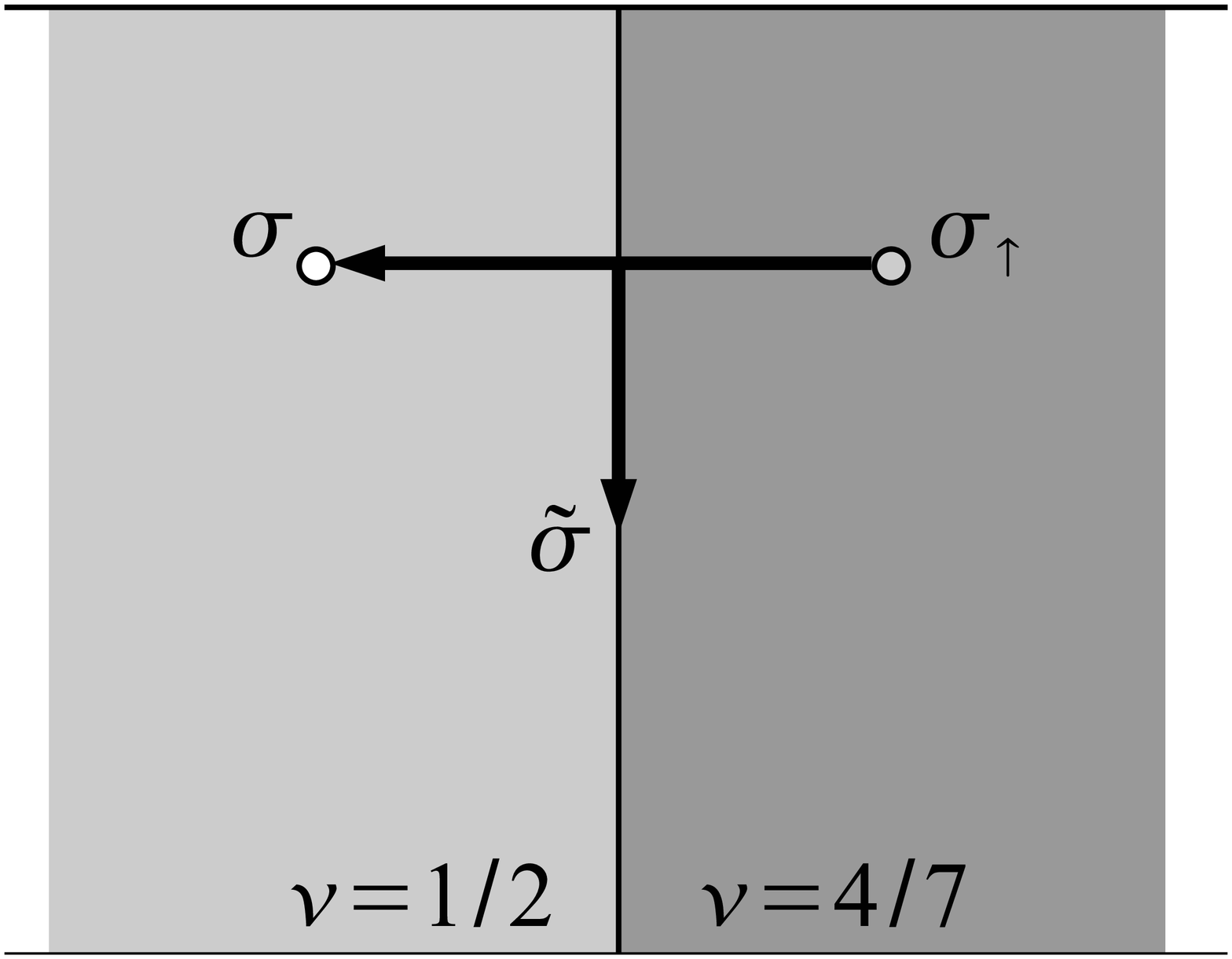}
      \includegraphics[width=0.45\textwidth]{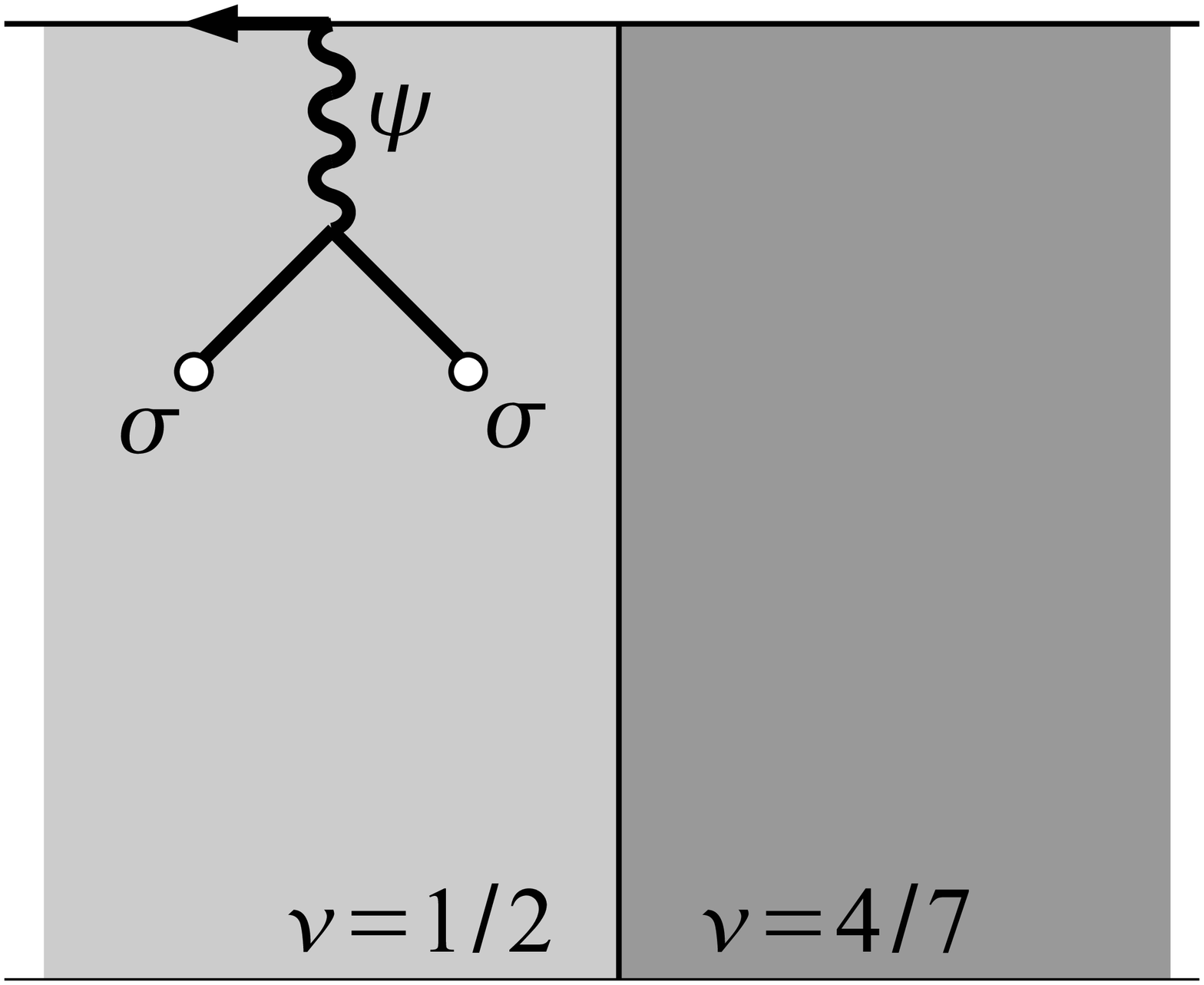}
    \end{center}
  \end{minipage}
  \hfill
  \vspace{0.1cm}
  \begin{minipage}[t]{0.5\textwidth}
    \begin{center}
      \includegraphics[width=0.45\textwidth]{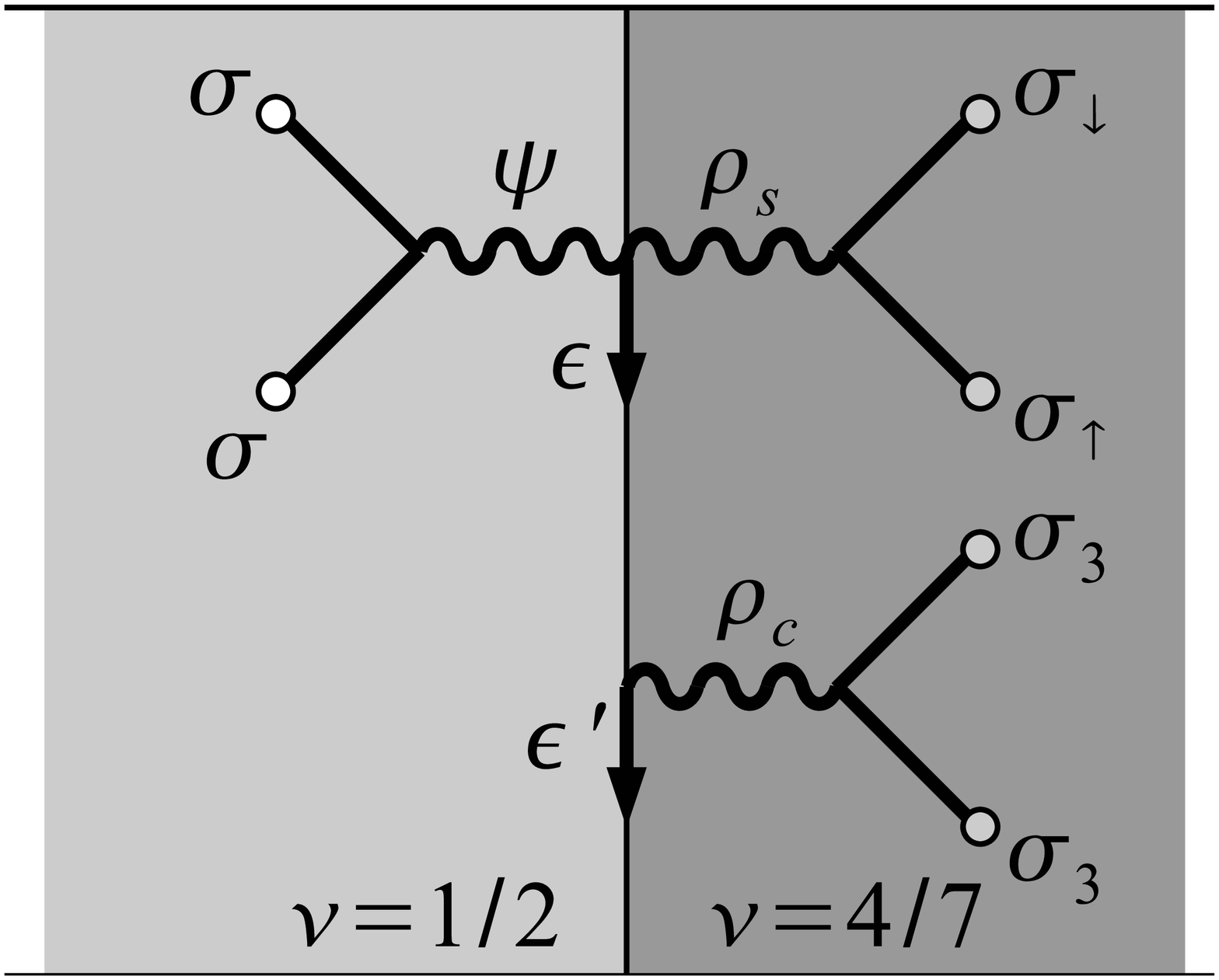}
      \includegraphics[width=0.45\textwidth]{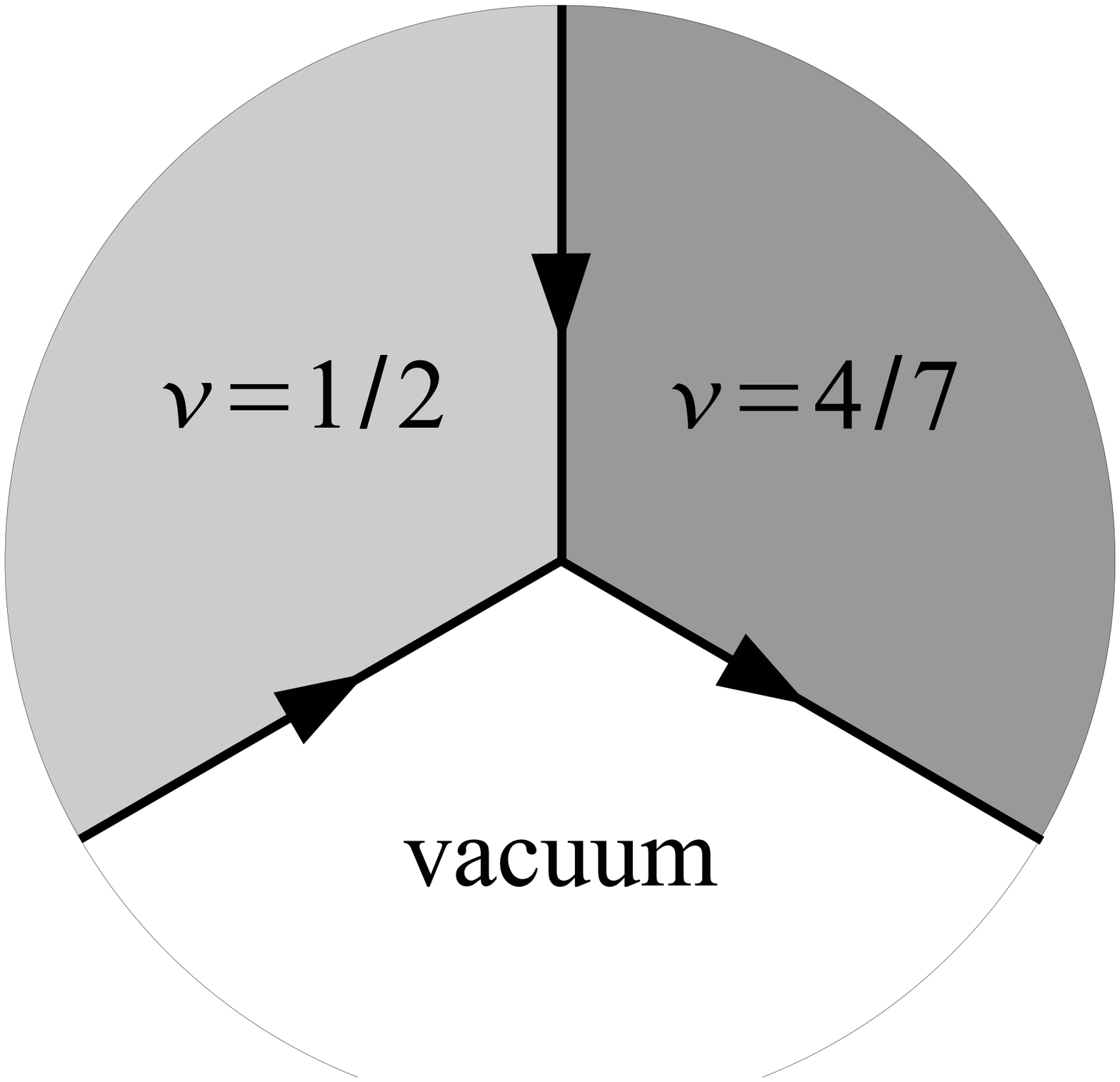}
    \end{center}
  \end{minipage}
  \hfill
  \caption{Upper-left: dragging a quasi-hole through the MR/NASS interface. 
           Upper-right: relaxation of an Ising qubit via edge to vacuum. 
           Lower-left: qubit relaxation via MR/NASS interface. 
           Lower-right: Y-junction.}
  \label{fig:edge}
\end{figure}

\paragraph{Gedanken experiments.}

To illustrate the role of the $c=7/10$ interface as a 'mediator' between two regions 
of different quantum statistical properties, we consider several Gedanken experiments.\newline
(i) \textit{Dragging a quasi-hole across the interface}. 
A $\sigma_\uparrow$-quasi-hole (charge $q=1/7$) is dragged 
from the $\nu=4/7$ side to the $\nu=1/2$ side, emerging as a $q=1/4$ 
$\sigma$ quasi-hole. During this process, a $\tilde{\sigma}$ is emitted 
into the edge, and a charge of $3/28$ absorbed. 
(Fig. \ref{fig:edge}, upper-left).\newline 
(ii) \textit{Qubit relaxation}. In both non-Abelian quantum Hall states, 
a pair of bulk quasi-holes constitute a qubit with two possible internal states. 
For a finite system, a qubit may relax its state by exchanging a 
neutral particle with a nearby edge or interface \cite{IGSS08},
through exponentially small tunneling matrix elements. 
An edge to vacuum can always relax a qubit state (see Fig. \ref{fig:edge}, 
upper-right, for an example where two quasi-holes in a MR state exchange 
a Majorana fermion with a MR/vacuum edge). 
Eqs. (\ref{eq:identifications}), (\ref{eq:identifications-rho}) lead to 
constraints on qubit relaxation via the MR/NASS interface. 
Two $\sigma_3$ quasi-holes can relax 
their state by exchanging a $\rho_c$ with a MR/NASS interface. However, this
same interface cannot relax the state of two $\sigma$ quasi-holes on the 
MR side, or of a $\sigma_\uparrow$-$\sigma_\downarrow$ qubit on the NASS side. 
The combined process, involving a $\rho_s$ on the NASS side and a $\psi$ on 
the MR side, is possible (Fig. \ref{fig:edge}, lower-left).\newline 
(iii) \textit{Y-junctions}. Consider a Y-juncion between an NASS state, a MR state 
and the vacuum (Fig. \ref{fig:edge}, lower-right). Thermal current 
through the junction splits proportionally to the central charge, unidirectionally 
as depicted in the figure, providing an observable which is directly sensitive to 
the gapping out of the two counter-propagating $c=1/2$ theories described in the 
above. 

Our discussion is easily generalized to an interface
between $k$-clustered RR and NASS states, leading to a neutral mode 
CFT of central charge $c_k=\frac{2(2k+3)(k-1)}{(k+3)(k+2)}$.

We close by mentioning some ideas related to the work here presented. 
The authors of \cite{BaisSlingerland08} discuss how condensing a boson can 
transform a non-Abelian topological phase (NA) into a phase with different 
topological order (NA'). This construction naturally leads to properties 
of an NA/NA' interface. In \cite{Gilsetal}, a finite density of non-Abelian anyons is shown to nucleate
a different topological liquid within a `parent' non-Abelian liquid. Their
interface is shown to provide examples of edge states between non-Abelian
phases. Clearly, the various approaches to NA/NA' interfaces are 
complementary, and they illustrate distinct features of the underlying 
physics.
 
We thank E.~Ardonne, F.A.~Bais and A.W.W.~Ludwig for discussions and for 
sending us their manuscripts prior to publication. This work was supported 
by the foundation FOM of the Netherlands, and by the ICMT.

\bibliographystyle{apsrev}
\bibliography{nass}

\end{document}